\documentclass[notitlepage,superscriptaddress,showpacs,nobalancelastpage,onecolumn,aps,longbibliography,prx,12pt]{revtex4-2}
\usepackage[english]{babel}
\RequirePackage[T1]{fontenc}
\RequirePackage{times} 

\usepackage[italicdiff]{physics}
\usepackage{siunitx} 
\usepackage{amsfonts}
\usepackage{lipsum}
\usepackage{subfigure}
\usepackage{amsmath}
\usepackage{amssymb}
\usepackage{wasysym}
\usepackage{xspace}
\usepackage{array}
\usepackage[hidelinks]{hyperref}
\usepackage[dvipsnames]{xcolor}
\usepackage{my_symbols}
\usepackage{CJK}
\usepackage{bm}
\usepackage{verbatim}
\usepackage{tikz}
\usepackage{marginnote}
\usepackage[textwidth=1.5cm]{todonotes}
\usepackage[normalem]{ulem}
\usepackage{soul}
\usepackage[commandnameprefix=ifneeded]{changes}

\sethlcolor{green}

{\par}

\newcounter{mycomment}

\begin{document}
\begin{CJK*}{UTF8}{gbsn} 
\title{Wave Computing based on Dynamical Networks: Applications in Optimization Problems}
\author{Yunwen Liu}
\affiliation{Department of Physics and State Key Laboratory of Surface Physics, Fudan University, Shanghai 200433, China}
\author{Jiang Xiao}
\email[Corresponding author:~]{xiaojiang@fudan.edu.cn}
\affiliation{Department of Physics and State Key Laboratory of Surface Physics, Fudan University, Shanghai 200433, China}
\affiliation{Institute for Nanoelectronics Devices and Quantum Computing, Fudan University, Shanghai 200433, China}
\affiliation{Shanghai Research Center for Quantum Sciences, Shanghai 201315, China}
\affiliation{Hefei National Laboratory, Hefei 230088, China}

\begin{abstract}

  We develop a computing framework that leverages wave propagation within an interconnected network, where nodes and edges possess wave manipulation capabilities, such as frequency mixing or time delay. 
  This computing paradigm can not only achieve intrinsic parallelism like existing works by the exploration of an exponential number of possibilities simultaneously with very small number of hardware units, but also extend this unique characteristic to a multidimensional space including spatial, temporal and frequency domains, making it particularly effective for addressing NP-hard problems.
   The proposed architecture has been validated through SPICE simulations, demonstrating its potential capability in solving several NP-hard problems, such as the Number Partitioning Problem, the 0/1 Knapsack Problem, and the Traveling Salesman Problem.

\end{abstract}

\maketitle
\end{CJK*}

\section{Introduction}

The classical computing paradigm, grounded in the Turing machine model \cite{turingComputableNumbersApplication1937} and the von Neumann architecture, has underpinned modern computing for over eighty years \cite{patterson_architecture_1990}. Nevertheless, this framework possesses inherent limitations, particularly regarding parallelism; enhancements in computational speed are predominantly achieved through the addition of cores or nodes, constrained by hardware scalability \cite{amdahl_validity_1967}. Such augmentations yield linear improvements, rendering them inefficient for addressing NP problems \cite{amdahl_validity_1967,gustafson_reevaluating_1988,karp_parallel_1990}. To address these shortcomings, various unconventional computing paradigms have emerged, including quantum computing \cite{nakamiya_qnn_2006}, probabilistic computing \cite{chowdhury_full-stack_2023}, and neuromorphic computing \cite{hopfield_nn_1982,hinton_bm_1985,mead_neuro_1990}.


As mentioned above, the efficient resolution of NP-hard problems is one of the primary challenges in computing \cite{karp_np_2010}, which includes combinatorial optimization problems such as the Traveling Salesman Problem and the Max-Cut Problem. These problems hold substantial relevance across various domains, including logistics, finance, and computer science. By definition, NP problems cannot be solved efficiently on deterministic Turing machines, yet they can be addressed in polynomial time on non-deterministic Turing machines \footnote{Weisstein, Eric W. Nondeterministic Turing Machine. From MathWorld -- A Wolfram Web Resource.}. The distinction between these two types of machines lies in the ability of the non-deterministic ones to explore multiple computational paths simultaneously. Currently, almost all classical computers operate as deterministic Turing machines, while non-deterministic Turing machines remain a theoretical construct used for analyzing computational complexity \cite{sengor_cnn_1998,gu_cnn_2004}.
The unfeasibility of non-deterministic Turing machines stems from the exponential parallelism they require. In comparison, the parallelism realized in conventional computing architectures scales linearly to the number of hardware units, such as CPU or GPU cores. 


In recent years, much attention has been drawn to novel computing schemes motivated by wave propagation, which basically rely on linear superposition law in real space \cite{wang_2021_inverse,zenbaa_2025_inverse}. Most of these works are focused on controlling the power flow \cite{khitun_ring_2022} or phase dynamics \cite{hoppensteadt_onn_1999} of oscillatory signals, 
where an array of oscillators are coupled through simple interconnecting edges \cite{csaba_coupled_2020} or oscillatory neural networks (ONNs) \cite{arai_spin_2018} 
to solve difficult problems such as combinatorial optimization \cite{balinskyy_tsp_2023,beattie_oscillator_2024} and machine learning \cite{aida_oscillator_2024}. In this work we emphasize another fundamental but rarely noticed characteristic of waves that facilitates inherent parallelism in the frequency domain. Specifically, the mixing or multiplication of two waves 
\begin{equation}
  \cos(\nu t)\cos(\omega t) 
  \propto \cos[(\nu-\omega)t] + \cos[(\omega+\nu)t]
\end{equation}
can be efficiently executed by a single device that accommodates all frequencies within the bandwidth. This mixing process splits a single peak at frequency $\nu$ into two distinct peaks at $\nu\pm \omega$ in the frequency domain. When an input signal comprises multiple frequencies $\nu_1, \nu_2,\dots,\nu_n$, this mixing operation allows for the simultaneous splitting of all $n$ frequencies. By integrating this wave-mixing functionality with the architecture of an interconnected network, exponential increase in the number of peaks can be realized. 
Such an unconventional way of wave manipulation has been utilized in previous research \cite{traversa_mem_2015}
, which turns out to have intrinsic parallelism and information overhead when dealing with some famous NP problems such as subset sum \cite{traversa_np_2015}, encoding problem specification into the frequency domain. 
Similar proposal also appears in neural network applications, such as multiplexed gradient descent that uses orthogonal oscillation in parameters for gradient estimation \cite{mccaughan_mgd_2023}. 
On this basis, we propose an more general wave computing framework based on dynamical networks capable of addressing certain additional NP problems, thereby demonstrating the potential of this approach in handling complex computational challenges.

This paper is organized as the following: Section \ref{sec:model} introduces the construction and the working principle of the wave-based dynamical network. Section \ref{sec:app} gives the examples of using the proposed computing scheme in solving three combinatorial NP problems: the Number Partitioning Problem, the 0/1 Knapsack problem, and the Traveling Salesman Problem. In Section \ref{sec:disc}, we discuss a few potential issues and future directions of this work.

\section{Dynamical Interconnected Network with Waves}
\label{sec:model}

\begin{figure*}[t]
  \includegraphics[width=\textwidth]{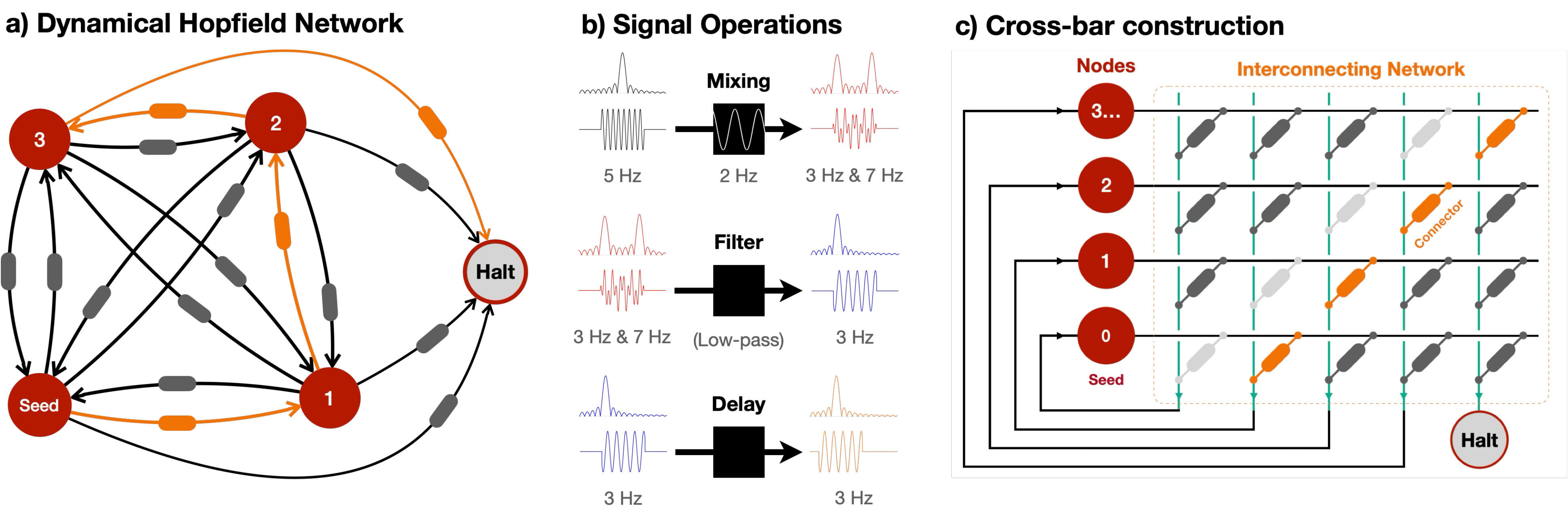}
  \caption{ 
  a) A graph representation of the dynamical network, where each node and edge is equipped with a functional device. The sub-graph connected by the orange edges illustrates a reduced network with a one-dimensional chain structure, which is used in the Number Partitioning Problem below.
  b) Each functional device may perform one or more operations in the frequency and time domain, including frequency mixing, shifting (not shown), filtering and time delay.  
  c) The construction of the Wave-based Dynamical Network using crossbar structure.
  }
  \label{fig:network}
\end{figure*}

To realize wave-based computing, a physical network 
can be designed as a directed complete graph as in \Figure{fig:network}(a) (with $N = 5$ nodes as an example), where each node or edge incorporates specific wave operation in frequency or temporal domain. These operations might be the frequency mixing, frequency filtering, or time delay operations as depicted in \Figure{fig:network}(b). A crossbar structure, shown in \Figure{fig:network}(c), provides an efficient layout for such a graph, with all nodes positioned on the left and connections implemented through an $N \times N$ array. 
The dynamical network comprises physical devices (nodes) interconnected by connectors (edges):
\begin{enumerate}
  \item The Seed node, which injects the initial signal into the network.
  \item The Intermediate node, which collects the signal from all incoming edges and re-emits the modified signal to all outgoing edges. 
  \item The Halt node, which has only incoming edges, monitors the arriving signal. If a signal with specific feature, such as one with a certain frequency, is detected, the computing procedure halts.  
  \item The Connector as an edge, which transmits signal from one node to another, executing frequency or time domain operation on the signal during this transmission. 
\end{enumerate}


Apart from the network topology, the most important ingredient in this new wave computing scheme is the functional elements located at the nodes and edges. In the following, we discuss a few possible frequency-domain and time-domain operations that can be placed in the nodes or on the edges.

\emph{Frequency Mixing -} 
The frequency mixing is to mix or, more specifically, multiply the incoming signal $I(t)$ with an internal signal in the node or on the edge. The simplest internal signal is the sinusoidal signal $\cos(\omega t)$, and thus the outgoing signal is 
\begin{equation}
  \label{eqn:mixing}
  O(t) = I(t) \cos(\omega t).
\end{equation}
Here the incoming signal $I(t)$ can be either continual waves or discrete wave packets, and the latter means the duration of the wave is small compared to the propagation time between nodes.
In frequency domain, the Fourier spectrum
\begin{equation}
  \label{eqn:fft}
  \td{O}(\nu)=\int O(\tau)e^{-i\nu\tau}\dd{\tau}.
\end{equation}
The mixing operation effectively splits all Fourier peaks of the incoming signal into two side peaks separated by twice of the characteristic frequency of the internal signal of the node or edge $2\omega$: 
\begin{equation}
  \label{eqn:mixing}
  \td{O}(\nu) \propto \td{I}(\nu+\omega) + \td{I}(\nu-\omega). 
\end{equation}
Therefore, as discussed in prior works \cite{traversa_mem_2015}, 
if the incoming signal $I(t)$ contains multiple peaks at $\qty{\nu_1, \nu_2, \cdots}$, the outgoing signal would carry twice as many peaks at $\qty{\nu_1 \pm \omega, \nu_2 \pm \omega, \cdots}$. The internal frequency $\omega$ can be different for each node or edge.

\emph{Frequency Shifting -} 
Another similar frequency domain operation is to shift the full spectrum of the incoming signal by a constant $\omega$ \cite{traversa_np_2015}:
\begin{equation}
  O(t) = I(t)e^{i\omega t}.
\end{equation}
In the frequency domain, this means
\begin{equation}
  \td{O}(\nu) = \td{I}(\nu + \omega).
\end{equation}
In many cases, the frequency mixing in \Eq{eqn:mixing} can also be used for the purpose of frequency shifting when only one of the side peaks plays the role and the other can be ignored.

\emph{Frequency Filtering -} 
A more complex operation is to non-locally modify the incoming signal in frequency domain with a frequency decoration function $\td{F}(\nu)$: 
\begin{equation}
  \td{O}(\nu) = \td{I}(\nu)\td{F}(\nu).
\end{equation}
In time domain, this is the convolution operation:
\begin{equation}
  O(t) = \int I(t-\tau) F(\tau)\dd{\tau}.
\end{equation}
Typical examples of the $\td{F}$ function include $\td{F}(\nu) = \Theta(\nu_M-\nu)$ for low-pass filter that only allows components lower than $\nu_M$ to pass, and $\td{F}(\nu) = \Theta(\nu - \nu_m)$ for high-pass filter that only allows components higher than $\nu_m$ to pass. They can be assembled to realize a band-pass filter $\td{F}(\nu) = \Theta(\nu_M-\nu)\Theta(\nu-\nu_m)$. Here $\Theta(x)$ is the Heaviside step function.

\emph{Time Delay -} 
A time-domain operation can be a delay by a certain amount $\tau$:
\begin{equation}
  O(t) = I(t-\tau) = e^{-\tau \dv{t}} I(t),
\end{equation}
where $e^{\tau\dv{t}}$ is the time translational operator. The time delay operation is important when the signals propagating in the network are wave packets which have finite duration in time.

\section{Applications}
\label{sec:app}

We now illustrate how to use this architecture to address some NP-hard problems, \ie the Number Partitioning Problem, the 0/1 Knapsack Problem, and the Traveling Salesman Problem. 

\subsection{Number Partitioning Problem (NPP)}

\emph{Problem Description - } Given a set of $N$ (positive) weights $\bw = \qty{w_0,...,w_{N-1}}$, find a partition into two disjoint subsets to minimize the total weight difference between the two subsets, \ie to minimize the absolute value of 
\[  D = \sum_j s_j w_j = \bs\cdot\bw 
\quad\mbox{with respect to}\quad
s_j = \pm 1.  \]

\begin{figure}[t]
  \includegraphics[width=\columnwidth]{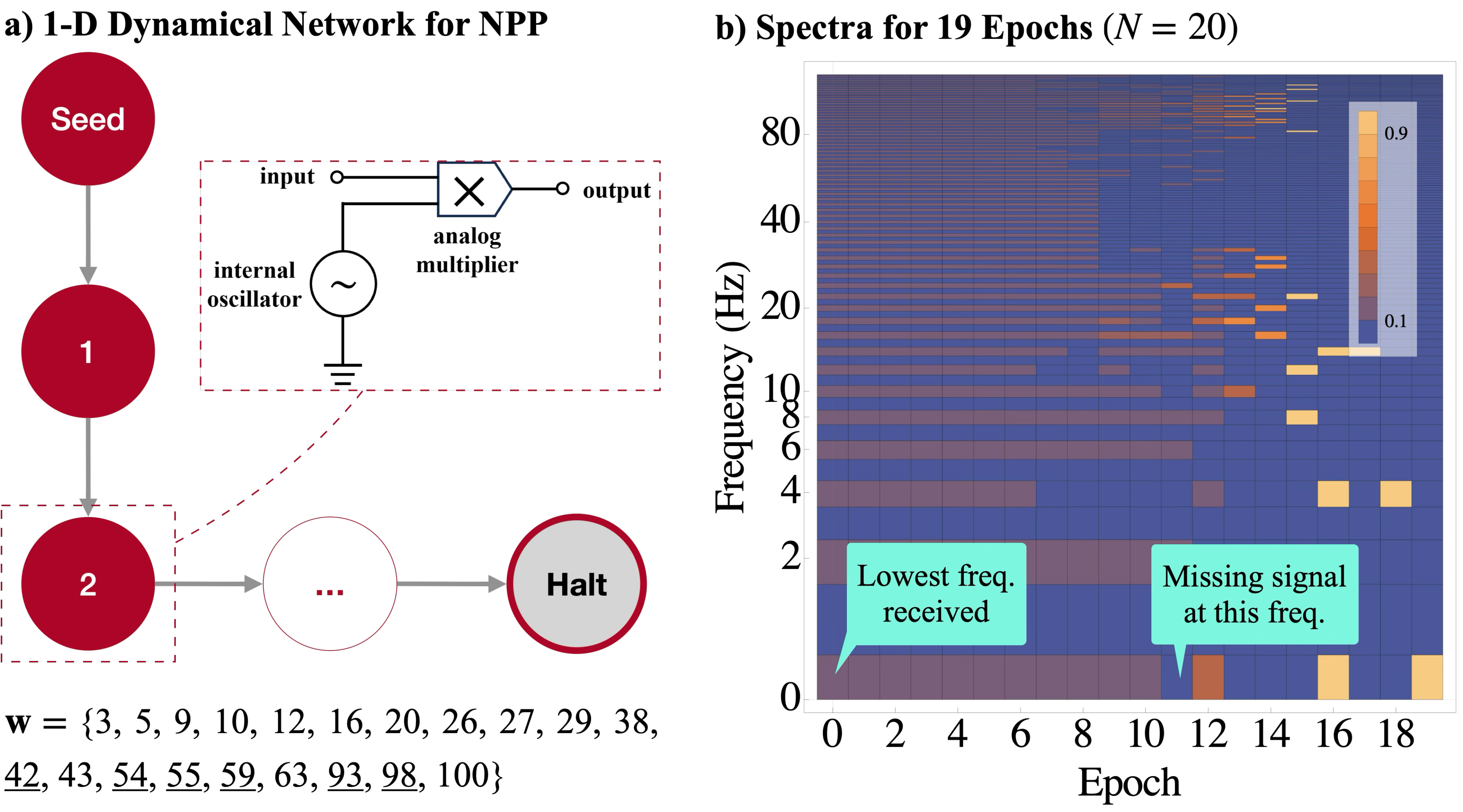}
  \caption{
  a) The schematic 1D chain for NPP. The inset shows the interior configuration of each node (mixer) in the 1D chain above, where the wave mixing operation is implemented by virtual analog multipliers. At the bottom is a $N=20$ example, encoded in the frequencies of the internal oscillators.
  b) The SPICE simulation result for NPP with $N = 20$ weights given in a).
  This problem has the minimal weight difference $D_m=0$ with the partition 
  $\qty{3,5,9,10,12,16,20,26,27,29,38,43,63,100}\cup\qty{42,54,55,59,93,98}$.
  The circuit is run for $N=20$ times to determine if each node belongs to the same subset as the seed node, and the FFT spectra of these epochs are shown here. The data has been renormalized and rescaled for better visualization. The existence of peaks at \SI{0}{Hz} shows that the $0\sim10,12,16,19$-th nodes should belong to the same subset for an optimal partition.
  }
  \label{fig:NPP}
\end{figure}

\emph{Network Construction for NPP}:
\begin{itemize} 
  \item The network for this problem is reduced from a complete graph to a 1D directed chain shown in \Figure{fig:NPP}(a), where the output of the $j$-th mixer is directly fed into the $(j+1)$-th mixer without modification;
  \item The $j$-th node performs a mixing operation by multiplying the incoming signal with its internal signal $\cos(\omega_j t)$ with the internal frequency given by the $j$-th weight: $\omega_j = w_j$; 
  \item The seed signal is a continual wave generated from the seed node at frequency $\omega_0$ and spread uni-directionally along the chain; 
  \item The lowest (non-negative) frequency received by the halt node at the end gives the smallest discrepancy among all possible weight partitions.
\end{itemize}

According to the network construction described above, the signal that reaches the halt node is a mixture of the characteristic signals from all $N$ nodes:
\begin{equation}
  \label{eqn:X_NPP}
  X(t) = \prod_{j=0}^{N-1} \cos(w_j t)
  \propto \prod_{j=0}^{N-1} \qty(e^{iw_j t} + e^{-iw_j t})
\end{equation}
From the expansion into sums, it is clear that the spectrum of signal $X(t)$ consists of frequencies given by $\bs\cdot\bw$ for all possible ($2^N$) choices of $\bs = \qty(s_0,\cdots,s_{N-1})\in\qty{-1,1}^N$ \cite{traversa_np_2015}:
\begin{equation}
  X(t) \propto \sum_{\bs} e^{it \bs\cdot\bw}. 
\end{equation}
Therefore, the minimum of the weight difference $D_m$ corresponds to the lowest (non-negative) frequency in the Fourier transform of $X(t)$.

Once the smallest weight difference \( D_m \) is determined, 
to deduce the exact partition, we refer to an incremental method by rerunning the process with some nodes merged into the seed node \cite{traversa_np_2015}. In the initial epoch, let node 1 be merged with the seed node, yielding \( w'_0 = w_0 + w_1 \). If the lowest frequency \( D_m \) remains observable in the resulting spectrum, it indicates that nodes \( w_1 \) and \( w_0 \) can indeed be put in the same subset, implying \( s_0 = s_1 = +1 \) (without loss of generality). Conversely, if the peak at \( D_m \) disappears from the spectrum, it implies that they should be in different subsets and \( s_1 = -1 \), prompting an adjustment to the merged seed as \( w'_0 = w_0 - w_1 \). This process can be extended to sequentially merging other nodes into the seed to determine their respective signs \( s_j \).

{\it SPICE Simulation - } We emulate the network using SPICE software ({\it NI Multisim 14.0}) to testify its validity and feasibility. The general chain topology is shown in \Figure{fig:NPP}(a) with $N = 20$ nodes. 
The design of each mixer node is in \Figure{fig:NPP}(b), 
in which the sinusoidal oscillator is denoted by an AC voltage source, which may be realized by some basic analog circuits such as the Wien bridge \cite{li_wien_2023} and the Colpitts oscillator \cite{azadmehr_colpitts_2020}. The mixing of two such signals is implemented by a virtual analog multiplier. The seed node contains only the oscillator and the output terminal. The halt node in \Figure{fig:NPP}(a) is emulated by a voltage probe, which records the waveform. Fourier transform is carried out by external numeric computation. 
The problem we try to solve here has $N = 20$ weights given in \Figure{fig:NPP}(a). 
The first run of the circuit gives a lowest frequency $D_m = \SI{0}{Hz}$ in the spectrum as seen in the first column (epoch 0) in \Figure{fig:NPP}(c), indicating that there is a partition with perfect balancing. 
The next task is to find the exact partition of the weights.
The circuit is then rerun for 19 more epochs with one weight merged into the seed node for each epoch. 
For example, in epoch 1, we skip node 1 and merge its weight into the seed node with $w_0^{(1)}=w_0+w_1=\SI{8}{Hz}$. The resultant spectrum still includes non-vanishing Fourier component at $D_m = \SI{0}{Hz}$, so we conclude that node 0 and node 1 belonging to the same subset ($s_0 = s_1 = +1$) may yield the optimal solution, and they are kept merged in the subsequent epochs. 
In epoch 2, we merge node 2 to the seed node and let $w_0^{(2)}=w_0^{(1)}+w_2=\SI{17}{Hz}$, and the appearance of Fourier peak at $D_m = \SI{0}{Hz}$ indicates that node 2 also belongs to the same partition as node 0 and node 1 ($s_2 = +1$). Similar merging processes are carried out for the rest of the nodes, and we found that node 1 - 10 all belong to the same partition as the seed node ($s_{0,\cdots, 10} = +1$). At epoch 11,
the seed node has $w_0^{(11)}=w_0^{(10)}+w_{11}=w_0+\cdots+w_{11}=\SI{237}{Hz}$, and the Fourier peak at $D_m=\SI{0}{Hz}$ disappears from the spectrum, indicating that node 11 does not belong the same partition as the seed node ($s_{11}=-1$).
After performing 19 epochs sweeping through all nodes, the exact partition is determined as indicated by the underlined numbers at the bottom of \Figure{fig:NPP}(a).
The example illustrated above turns out to have a perfect partition with $D_m = 0$. Two examples that have imperfect partitions with $D_m > 0$ are included in the Supplemental Material 
\cite{sm}
with different weight ranges (for $N=8$ and $N=12$).

\subsection{0/1 Knapsack Problem (0/1-KP)}

\emph{Problem Description -} Given a set of $N$ items with weights $\bw = \qty{w_1,\dots,w_N}$ and values $\bv = \qty{v_1, \dots, v_N}$, find a subset (item combination) with maximal total value without exceeding a given weight capacity $W$ of the knapsack, \ie find a set of $\qty{x_j}$ to maximize 
\[  V = \sum_{j=1}^N x_j v_j 
\qwith \sum_{j=1}^N x_j w_j \le W
\qand x_j = 0, 1.
\]




In the previous example of NPP, the signal is a continual wave, with frequency discernible only at the halt node. However, by introducing an additional temporal degree of freedom, the pulsed signals (wave packets) travel across the network in both frequency and time domain. The halt node monitors both the arrival time and the spectrum of the incoming signal. In the following network for 0/1-KP, we use the frequency to encode the value, and the delay/arrival time to encode the weight.

\begin{figure*}[t]
  \includegraphics[width=\textwidth]{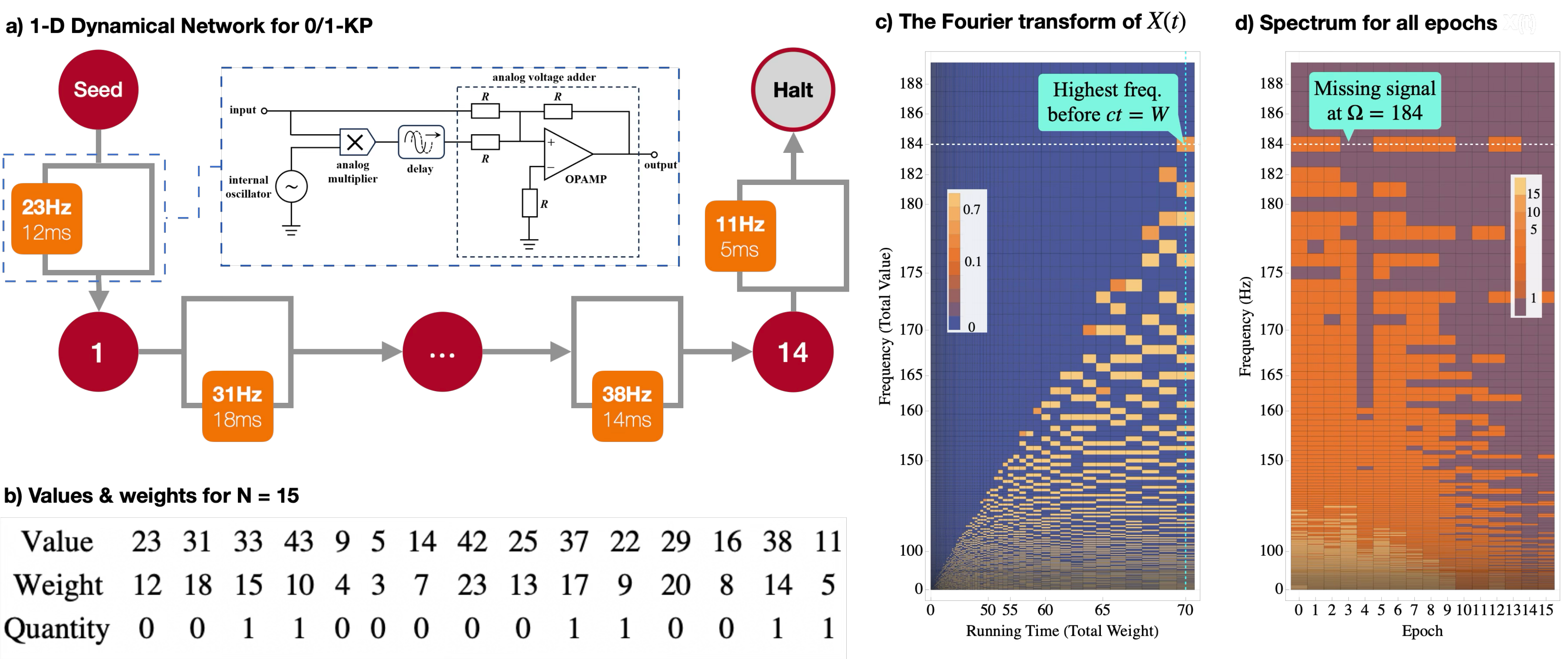}
  \caption{
  a) The construction of dynamical network for the 0/1-KP. In the inset is the interior configuration of each edge in the 1D chain above. A wave packet will either pass through with no change via the upper wire, or be mixed with the sinusoidal signal followed by a delay via the lower branch. The signals from the two branches are summed up by the OPAMP before transmitting downstream.
  b) The example problem we solve in the main text. The `Quantity' row shows that the optimal solution is to take the 3,4,10,11,14,15-th items in the knapsack, which can be verified by traditional methods.
  c) The temporal-frequency distribution of wave packets in the 0-th epoch (when all the edges are included) for the example problem in b). The colors stand for (rescaled) amplitudes of Fourier components.
  The dashed lines highlight the signal with the highest frequency (value) before time $ct = W = 70$ (the capacity of the knapsack), which happens at $ct = 70\le W$ and has value $V = 184$. 
  d) The counts of hitting different frequencies in each epoch. The white dashed line shows the maximal total value $184$. In the 3,4,10,11,14,15-th epochs, no wave packets hit this value, so the corresponding items should belong to the optimal solution.
  }
  \label{fig:KP}
\end{figure*}

\emph{Network Construction for 0/1-KP}:
\begin{itemize} 
  \item The network has the 1-dimensional chain-like structure as shown in \Figure{fig:KP}(a).

  \item All nodes do nothing to the signal except transmission downstream.

  \item The edge connecting the $j$-th node to the $(j+1)$-th node has two branches, one doing nothing, and the other performing a frequency shift (or mixing) operation by $v_j$ and a time delay operation by $w_j$. The combined signal is fed into the $(j+1)$-th node.

  \item The seed signal is a wave packet (of zero central frequency) with short duration $\delta$, \eg a Gaussian wave
  \[ X_0(t) = e^{-t^2/\delta^2} \]
  emitted from the seed node and spread along the chain.

  \item The halt node watches all incoming wave packets until time $t = W$ (the knapsack capacity), and the wave packet with the highest frequency represents the maximal total value constrained by the knapsack capacity.
\end{itemize}

According to the network construction described above, the time series of the signal reaching the halt node can be expressed as
\begin{equation}
  \label{eqn:X_KP}
  X(t) = \prod_{j=1}^N \qty(1+ e^{iv_j t}e^{-w_j \dv{t}}) X_0(t),
\end{equation}
where the two terms in the bracket represent the two branches of the $j$-th edge. Specifically, the unity term represents the trivial branch of doing nothing, and other term represents the branch with a time translation $e^{-w_j\dv{t}}$ followed by a frequency shift $e^{iv_jt}$ (or a frequency mixing with $\cos(v_jt)$). 
When \Eq{eqn:X_KP} is expanded, it is the sum of ($2^N$) wavepackets with frequency $\bx\cdot\bv$ and delay $\bx\cdot\bw$ for all possible choices of $\bx = (x_1,\cdots,x_N)\in\qty{0,1}^N$:
\begin{equation}
  X(t) \propto \sum_{\bx} e^{i \bx\cdot\bv(t-\bx\cdot\bw)} X_0(t-\bx\cdot\bw) 
\end{equation}
Therefore, the signals arriving at the halt node correspond to all possible item combinations with the total value encoded in the frequency of the wavepacket and the total weight encoded in the arrival time.  
The Fourier components of the signal arriving at the halt node at different time instants $t$ can show the spectra of the wave packets:
\begin{equation}
  \label{eqn:Xtw}
  \td{X}(\nu, t) = \int_{t}^{t+\delta} X(\tau) e^{-i\nu\tau}\dd{\tau}.
\end{equation} 
The solution for the 0/1-KP corresponds to the peak in $\td{X}$ with highest frequency $\nu$ received before time $t = W$ (the knapsack capacity). Once the optimal total value is identified, the exact combination can be determined by rerunning the network assuming one item is excluded from the knapsack. If the rerun yields the same optimal total value as the original run, then this particular item should indeed be excluded, otherwise it should be included. By excluding items one by one, a specific combination can be inferred within approximately $N$ reruns.


\emph{SPICE Simulation - } We demonstrate the process using SPICE simulation. 
The seed node is implemented with a voltage source that generates a square wave packet with duration $\delta=1$ (a simplification to the Gaussian packet), and the other nodes are trivial. The edges with two branches are constructed as in the inset of \Figure{fig:KP}(a), 
where the input signal can be either transmitted via the line on the top without modification, or via the lower line with frequency mixing (by the multiplier) and time delay (by the virtual `Delay' component in SPICE). 
These two channels of signals are summed up by a sub-threshold Operational Amplifier (OPAMP). 
The time delay on the lower channel is set as proportional to the corresponding weight: \(\tau_j = w_j/c\), where \(c\) is equivalent to the wave velocity. 
Given that the weights take integer values, the minimal time separation between the arrival times of distinct wavepackets is \(1/c\).
Since the duration of each wave packet is \(\delta = 1\), setting \(1/c=2\) can avoid the overlapping of signals with different arrival times.

Here we show the results for a problem ($N=15$) with weights and values listed in \Figure{fig:KP}(b). The SPICE simulation results for two other problems ($N = 8$ and $20$) are shown in the Supplemental Material \cite{sm}. 
The Fourier spectrum of the signal received at the Halt node at all times ($\td{X}(\nu,t)$) is plotted in \Figure{fig:KP}(c), which shows the peak with maximal frequency happens at $\nu = 184$ with delay $ct = 70$ ($\le W = 70$), matching the exact solution of the problem by pure enumeration. 
Because we used the frequency mixing rather than frequency shifting in the SPICE simulation,
there are additional frequency peaks in \Figure{fig:KP}(c) that do not belong to the possible values of $\sum_{j=1}^N x_j v_j$ (with $x_j=0,1$), but these additional signals always have lower frequency (thus lower total value) and do not influence the highest frequency peak. 
After finding the optimal total value (184) with the initial run (epoch 0), we rerun the network for $N=15$ more times to determine the exact combination of items. 
Since the exact arrival time, as long as no later than $t=W/c$, is not of our concern, we may accumulate all the signals during time $t\in [0,W/c]$. And the optimal solution is marked by the highest frequency hit in the accumulated spectrum shown in \Figure{fig:KP}(d). 
In epoch 1, node 1 is skipped, and the wave packets can still hit $\nu = 184$ before time $t = W/c$, so item 1 shall not be in the knapsack and is excluded in the subsequent epochs. The same result holds for item 2. In epoch 3, when node 3 is skipped, no wave packet hits the frequency of $184$ before $t=W/c$. Therefore, item 3 should be included in the knapsack for the optimal combination, and in the following epochs it is kept in the network all the time. By successively skipping the rest of the nodes in the following epochs, we found that the optimal signal at $184$ is absent when items $3,4,10,11,14,15$ are skipped, indicating that they shall be in the knapsack, and the other items shall not. This also agrees with the correct answer $\bx=(0,0,1,1,0,0,0,0,0,1,1,0,0,1,1)$ by exact enumeration.

\subsection{Traveling Salesman Problem (TSP)}

\emph{Problem Description -} Given a $N\times N$ distance matrix $L$ for $N$ cities $0,1,\cdots,N-1$ with $L_{jk}$ as the distance between city $j$ and $k$, find the shortest Hamiltonian cycle for a salesman who must visit each city exactly once and finally return to the starting city.\\


\emph{Network Construction for TSP}:
\begin{itemize}
  \item The network is a complete (undirected) graph  with $N$ nodes as in \Figure{fig:network}(a).

  \item All non-seed nodes perform frequency shifting operation by a node-dependent characteristic frequency $\omega_j$. 

  \item All edges are equipped with time-delay operation with delay time proportional to the corresponding distance, \ie $\tau_{jk}=L_{jk}/c$; 

  \item Unless otherwise stated, the shifted signals emitted from each node will be transmitted along all the edges connected to it.


  \item The seed node emits an initial wave packet with central frequency $\omega_0$ at time $t = 0$: 
  \[ X_0(t) = e^{i\omega_0 t} e^{-t^2/\delta^2}.\]
  After emitting the initial weave, the seed node becomes the halt node. 

  \item The halt (seed) node monitors the spectra of the returning wave packets and watches if the spectra match presumed features, such as if a signal with a target frequency $\Omega$ appears. 
\end{itemize}


According to the network construction described above, the signal that returns to the halt (seed) node can be expressed as
\begin{equation}
  \label{eqn:X_TSP}
  X(t) = \sum_\gamma \prod_{e_{jk}\in\gamma} e^{i\omega_j t}e^{-\tau_{jk} \dv{t}} X_0(t),
\end{equation}
where $\gamma$ represents all closed paths that start and end at the seed/halt node, and $e_{jk}$ denotes the edge from node $j$ to $k$. 
\Eq{eqn:X_TSP} includes all Hamiltonian loops, as well as many non-Hamiltonian loops. The signal that returns to the seed/halt node via path $\gamma$ is characterized by its arrival time and its frequency:
\begin{equation}
T_\gamma = \frac{1}{c}\sum_{e_{jk}\in \gamma} {L_{jk}} \qand
\Omega_\gamma = \sum_{j\in\gamma} \omega_j.
\end{equation}
We should note that if the path $\gamma$ is non-Hamiltonian, the same edge or node may appear multiple times in $\gamma$, and each appearance should be summed over independently.

\begin{figure*}[t]
  \includegraphics[width=\textwidth]{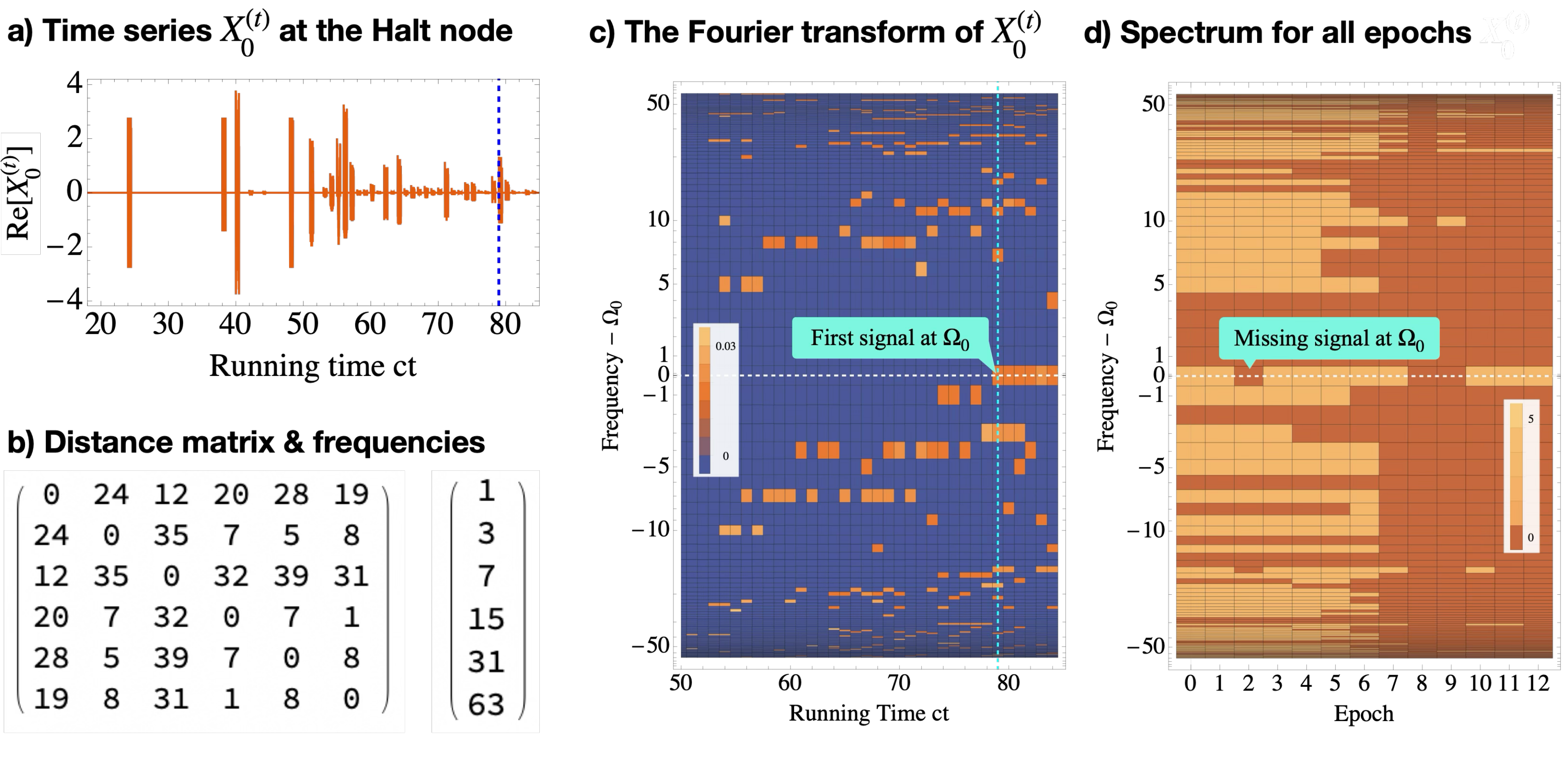}
  \caption{FIFO simulation result of an example of 6-city TSP. 
  a) The wave packets in time domain reaching the seed/halt node as function of time $X_0^{(t)}$. 
  b) The distance matrix (left) and the allocation of characteristic frequencies for each node (right). The shortest Hamiltonian cycle has distance $d_0 = 79$.
  c) The wave packets in frequency domain as function of time. Each wave packet in a) contains multiple frequency components. The first signal with sum frequency $\Omega_0$ returns at $ct_0 = 79$, marking the shortest Hamiltonian cycle.
  d) Once the length of the shortest cycle is found, we rerun the network to determine the edges that do not belong to the cycle.
  }
  \label{fig:TSP_FIFO}
\end{figure*}

For a Hamiltonian loop, each city is visited exactly once, and thus the corresponding signal is stamped with the sum-frequency: 
\begin{equation}
  \label{eqn:Omega_TSP}
  \Omega_0 = \sum_{j=0}^{N-1}\omega_j.
\end{equation}
If we choose the characteristic frequencies $\qty{\omega_j}_{j=0}^{N-1}$ properly such that \Eq{eqn:Omega_TSP} is the only way to yield the sum-frequency, we can use the appearance of signal with frequency $\Omega=\Omega_0$ as an indicator for a Hamiltonian cycle. Other closed but non-Hamiltonian loops shall have different frequencies and can be screened out. With this condition, the first signal returning to the seed node with sum-frequency $\Omega_0$ represents the shortest Hamiltonian cycle, or the solution to the TSP. 

The exact shortest Hamiltonian cycle can be determined by rerunning the network with edges removed one by one. If it takes more time to detect a sum-frequency signal, the removed edge belongs to the shortest Hamiltonian cycle, otherwise not. In this way, the shortest Hamiltonian cycle can be determined within approximately $N^2$ runs.

\emph{Numerical Simulation with FIFO - } 
Because the network for TSP is a complete graph with $\sim N^2$ edges and works on complex numbers, the SPICE simulation becomes impractical. Instead, we utilize the data structure of First-In-First-Out (FIFO) to simulate the real time wave packet propagation in the network. The detailed FIFO based algorithm is described in the Supplemental Material \cite{sm}. 

The main idea is to use FIFO as pipelines with sufficient number of blocks to represent the real time waves on all edges. To be more specific, let the pipeline representing the edge connecting node $j$ and node $k$ at discrete time $t$ to be $P_{jk}^{(t)}(l)$, where $l = 0, 1, \cdots, L_{jk}/c$ represents the location along the length from node $j$ to node $k$. The total number of elements is given by the total length of the map: $\sum_{j<k} L_{jk}/c$. The pipelines keep updating themselves $P^{(t)} \mapsto P^{(t+1)}$, mimicking the propagation of waves in the network, according the following rule:
\begin{itemize}
  \item At $t = 0$, initialize all pipelines with zeros: $P^{(0)}_{jk}(l) = 0$ with $l = 0, 1, \cdots, n_{jk} = L_{jk}/c$.

  \item The seed node starts to emit a square wave packet of duration $\delta$ by setting $P_{0k}^{(t)}(0) = e^{i\omega_0 t} \Theta(\delta - t)$ for all $k\neq 0$. 

  \item For each time step, the data shifts one position along the pipeline: $P_{jk}^{(t+1)}(l+1) = P_{jk}^{(t)}(l)$ for all $0 \le l < n_{jk}$. 

  \item Popping out the last item in each pipeline and injecting their average as the input signal to the connected node: $X_k^{(t)} = N^{-1}\sum_{j\neq k} P_{jk}^{(t)}(n_{jk})$.

  \item Mixing the input signal to each node with the node-dependent internal frequency, and popping in the mixed signal into the beginning of all connecting pipelines: $P_{kl}^{(t)}(0) = X_k^{(t)} e^{i\omega_k t}$. 

  \item Recording the popped out signal at the halt/seed node $X_0^{(t)}$.
\end{itemize}

Since there are efficient classical algorithms (such as the greedy estimation of the nearest neighbor (NN) algorithm \cite{2019Halim}) that can determine the upper bound $L_g=cT_g$ of the shortest Hamiltonian loop, this wave approach is guaranteed to yield a valid solution before time $T_g$. Therefore, the updating process does not have to run infinitely.

\Figure{fig:TSP_FIFO} shows the results of the FIFO simulation for an illustrative problem with $N=6$ cities 
\footnote{This example is a non-Euclidean TSP, so the distance matrix does not satisfy triangular inequality.}.
The results for a slightly larger problem ($N=8$) is included in the Supplemental Material \cite{sm}. 
In the $N=6$ problem, the total running time is indicated by $L_g = cT_g = 84$ obtained from a greedy estimation of the nearest neighbor (NN) algorithm \cite{2019Halim}. 
\Figure{fig:TSP_FIFO}(a) shows the temporal sequence $X_0^{(t)}$ received at the halt/seed node as function of time, and \Figure{fig:TSP_FIFO}(c) shows its moving Fourier transform (as defined in \Eq{eqn:Xtw}) with the sum-frequency $\Omega_0 = 120$ subtracted.
It is observed that the first signal at the sum-frequency $\Omega_0$ occurs at running time $ct_0 =79$, 
which corresponds exactly to the optimal solution for length of the shortest Hamiltonian cycle for this particular problem ($d_0 = 79$). Therefore, this demonstrates that the arrival time of the first signal with sum-frequency returning to the seed node yields the solution of the TSP. 



Once the length of the shortest Hamiltonian cycle is found in the first run (epoch 0) of the network, we continue to rerun the network with edges removed one by one to determine the exact cycle. Similar to the 0/1-KP, in each of the subsequent epochs we remove one edge from the graph by setting its length to be $\infty$. Since the wave packets with frequency $\Omega_0$ will not reach the halt node earlier than $t_0$, in each epoch we mark the existence of the original optimal loop by the number of hitting $\Omega_0$ before time $t_0$ in the accumulated spectrum in \Figure{fig:TSP_FIFO}(d). If there is no hitting at $\Omega_0$, then the removed edge should belong to the shortest Hamiltonian cycle, and this edge is put back into the network. On the other hand, if the wave packet at $\Omega_0$ remains, then the removed edge is non-essential and will be excluded from all the following epochs. At most $N(N-1)/2$ epochs are required to determine the exact Hamiltonian cycle, and some further optimization strategies can be adopted to reduce the number of epochs. For instance, the longer edges can be checked earlier because they are more likely to be excluded. Furthermore, there is no need to check the last two edges of any nodes because they must be included to form a Hamiltonian cycle. The edge removing process can stop when there are $N$ edges left.
For the 6-city problem in \Figure{fig:TSP_FIFO}, we first remove the longest edge $L_{24}=39$ in epoch 1, for which the wave packet at $\Omega_0=120$ remains, so $L_{24}$ can indeed be excluded. In the next epoch, we try to remove the second-longest edge $L_{12}=35$, and the target wave packet disappears, which means $L_{12}$ should be included. By continuing this removing process on the shorter edges, we successfully removed 9 edges in 12 epochs (less than $6\times 5/2 = 15$ epochs), and the remaining 8 edges form the shortest Hamiltonian cycle: (0214350).

\section{Discussion}
\label{sec:disc}

The maximal computational power of such a dynamical network can be achieved only when it is constructed physically based on hardware that can operate with wide bandwidth. On top of its inherent parallelism in frequency domain \cite{traversa_np_2015}, 
we develope this key advantage of the dynamical network across multiple dimensions - namely, the frequency, spatial, and time domains. Specifically, signals of varying frequencies can propagate simultaneously within the network's space-time framework. This multidimensional parallelism is directly proportional to the total volume of the multidimensional space, 
\ie the product of the number of nodes/edges, the total running time, and most importantly the bandwidth of the frequency spectrum. The feature that the parallelism is proportional to the frequency bandwidth, not simply the number of hardware units, is a unique advantage of this wave-based network. For this reason, the new computing scheme has potentially surpassed the parallelism offered by conventional supercomputers, whose parallelism linearly depends on the number of cores.

Even though this wave-based computing scheme has the potential to solve NP problems efficiently, we suspect the realized calculation falls in the scope of pseudo-polynomial complexity \cite{traversa_mem_2015,bertsekas_dp_2017}, \ie it is efficient when the (sum of the) numbers involved in the NP problem do not increase exponentially as function of the size of the problem. This pseudo-polynomial property definitely holds for NPP and 0/1 KP, but it may encounter some trouble in TSP. To faithfully distinguish Hamiltonian cycles from other loops, the target frequency $\Omega_0$ that the halt/seed node keeps watching must be unique, \ie the spectrum for a non-Hamiltonian cycle shall not include $\Omega_0$. In order to realize this, a careful choice of frequency set $\qty{\omega_j}$ \cite{2000Nathanson,2018Zhou}, or cross-check using multiple frequency sets, is required.

The energy density within a network can also pose challenges. In the context of the NPP, the seed signal is concentrated at a single frequency; however, as it traverses the network, this energy disperses, divided into two frequencies after the first node and subsequently into four after the second, and so forth. This process diminishes the power of each frequency peak, potentially leading to a critically low signal-to-noise ratio after multiple nodes have been traversed. To mitigate this issue, it becomes essential for each node to amplify the signal. However, excessive amplification can result in divergent total energy. Consequently, meticulous control of the power density across all frequency peaks might be critical. This necessity for power regulation similarly extends to 0/1-KP and TSP.

This paper focuses on developing the theoretical framework of a novel wave-based computing scheme, demonstrating its potential through simulations that address NP-hard problems of limited scale. To fully harness the capabilities of this approach, it is imperative to develop a physical network adept at manipulating high-frequency waves, such as microwaves in the GHz range or optical waves at THz frequencies. Additionally, achieving a wide bandwidth is crucial to enhance parallel processing capabilities, thereby maximizing the efficiency and effectiveness of this computing paradigm.

\section{Conclusions}

On top of the existing paradigm utilizing the natural parallelism of propagating waves, we propose a more versatile wave-based dynamical network, which can make versatile operations in space, time and frequency domain to encode information in optimization problems. The performance of such a network is derived mathematically and then realized numerically in NPP, 0/1 KP and TSP, indicating that its computational capacity is not constrained by the number of hardware cores, but related with the volume used in the multidimensional state space. However, the present scheme still encounters the problem of vanishing energy density, and is likely to fall into the regime of pseudo-polynomial methods. 
Diverse investigation can be carried out in the future to study its practicality and improve its efficiency on algorithm and architecture.\\


\bigskip

\emph{Acknowledgements - } 
This work was supported by 
National Natural Science Foundation of China (Grants No. 12474110),
the National Key Research and Development Program of China (Grant No. 2022YFA1403300),
the Innovation Program for Quantum Science and Technology (Grant No.2024ZD0300103),
and Shanghai Municipal Science and Technology Major Project (Grant No.2019SHZDZX01).\\

The data that support the findings of this article (including the Supplemental Material) are openly available \cite{data_NPP,data_KP,data_TSP}.

\bibliography{wavebib, DHN}

\end{document}